\documentstyle[prl,multicol,aps,epsf]{revtex}
\tighten
\begin{document}

\title{Observation of Bound Surface States in Grain Boundary Junctions of
High Temperature Superconductors}

\author{L. Alff$^1$, A. Beck$^2$, R. Gross$^1$, A. Marx$^1$, S. Kleefisch$^1$,
Th. Bauch$^1$, H. Sato$^3$, M. Naito$^3$, and G. Koren$^4$}

\address{$^1$II.~Physikalisches Institut, Universit\"{a}t zu K\"{o}ln,
Z\"{u}lpicherstr.~77, D - 50937 K\"{o}ln, Germany}

\address{$^2$IBM Research Division, Z\"{u}rich Research Laboratory,
S\"{a}umerstr. 4, 8803 R\"{u}schlikon, Switzerland}

\address{$^3$NTT Basic Research Laboratories, 3-1 Morinosato Wakamiya,
Atsugi-shi, Kanagawa 243, Japan}

\address{$^4$Physics Department, Technion, Israel Institute of Technology,
32000 Haifa, Israel}

\date{received February 10, 1998}
\maketitle

\begin{abstract}
We have performed a detailed study of the tunneling spectra of bicrystal grain
boundary junctions (GBJs) fabricated from the high temperature superconductors
(HTS) YBa$_2$Cu$_3$O$_{7-\delta}$ (YBCO), Bi$_2$Sr$_2$CaCu$_2$O$_{8+\delta}$
(BSCCO), La$_{1.85}$Sr$_{0.15}$CuO$_4$ (LSCO) and
Nd$_{1.85}$Ce$_{0.15}$CuO$_{4-y}$ (NCCO). In all experiments the tunneling
direction was along the CuO$_2$ planes. With the exception of NCCO, for all
materials a pronounced zero bias conductance peak (ZBCP) was observed which
decreases with increasing temperature and disappears at the critical
temperature. These results can be explained by the presence of a dominating
$d$-wave symmetry of the order parameter resulting in the formation of zero
energy Andreev bound states at surfaces and interfaces of HTS. The absence of
a ZBCP for NCCO is consistent with a dominating $s$-wave symmetry of the pair
potential in this material. The observed nonlinear shift of spectral weight to
finite energies by applying a magnetic field is in qualitative agreement with
recent theoretical predictions.
\end{abstract}


\vspace*{-10cm}\noindent
To appear in Physical Review B
\vspace*{9.5cm}

\vspace*{-1.25cm}
\begin{multicols}{2}
\narrowtext

There is strong evidence that the superconducting order parameter (OP) in the
HTS has a dominating $d$-wave symmetry \cite{vHarlingen:95,Scalapino:95}. For
this pairing symmetry there is a $\pi$-phase shift of the OP in orthogonal
$k$-space directions resulting in a positive and negative sign of the pair
potential in those directions. This also means that there are directions with
nodes of the pair potential, e.~g.~for a pure $d_{x^2-y^2}$-symmetry, the
nodes are along the [110] direction in the CuO$_2$ plane. For the tunneling
spectra of junctions employing HTS electrode materials with a $d$-wave
symmetry of the OP, a pronounced ZBCP has been predicted originating from
mid-gap surface (interface) states or zero energy bound states (ZES) at the
Fermi level
\cite{Hu:94,Tanaka:95b,Barash:95a,Buchholtz:95,Kashiwaya:96a,Hu:98}. The
physical reason for these states originates from the fact that quasiparticles
incident and reflecting from the surface propagate through different order
parameter fields which leads to Andreev reflection. The constructive
interference between incident and Andreev reflected quasiparticles results in
bound states. Stable ZES are formed if the scattering induces a change in sign
of the OP. For a $d_{x^2-y^2}$-wave symmetry such sign change and, hence, the
presence of ZES is possible for all surfaces parallel to the $c$-axis except
for those with the lobe directions perpendicular to the surface, whereas for a
$s$-wave symmetry no ZES are possible. The spectral weight of the ZES for a
$d_{x^2-y^2}$-wave symmetry depends on the orientation of the surface with
respect to the crystal axis. The maximum spectral weigth is expected for a
(110) surface and, hence, a maximum ZBCP is expected for tunneling in the
direction of the nodal lines, i.\ e.,\ the [110] direction. This has been
observed recently using low temperature scanning tunneling spectroscopy
(LTSTS) \cite{Alff:97b} and planar type junctions \cite{Covington:97}. We note
that the ZBCP is sensitive to surface roughness making it difficult to
distinguish between the directions in the plane
\cite{Matsumoto:95a,Yamada:96,Barash:97}.

Initially, the ZBCP in the tunneling spectra of HTS junctions has been
explained within the Appelbaum-Anderson (AA) model \cite{Appelbaum:66a} due to
the presence of a large density of magnetic scattering centers at the surface
of the junction electrodes. However, the AA-model predicts a ZBCP that is not
expected to disappear at a certain temperature and to split {\em linearly}
with increasing applied magnetic field. Furthermore, it has been suggested
recently that the surface of $d$-wave superconductors might show spontaneously
generated surface currents \cite{Matsumoto:95a,Sigrist:95a} and a phase with
broken time-reversal symmetry \cite{Fogelstroem:97a}. In such state the
Andreev bound states shift to finite energies resulting in a splitting of the
ZBCP even in zero magnetic field \cite{Fogelstroem:97a}. Applying a magnetic
field results in a further splitting of the ZBCP. However, in contrast to the
AA-model prediction this splitting is predicted to increase {\em non}linearly
with applied field. In order to clarify these issues experimentally and to
rule out competing explanations for the origin of the ZBCP in GBJs, in this
report we present a comprehensive analysis of the ZBCP for different materials
including YBCO (60\ K-phase and 90\ K-phase), BSCCO, LSCO, and NCCO. We
emphasize that for YBCO, BSCCO and LSCO the OP is considered to have a
dominating $d$-wave component \cite{Achsaf:96}, whereas for NCCO the
dominating component most likely is a $s$-wave as suggested by several
experiments \cite{Huang:90,Wu:93,Andreone:94,Alff:96b}. Therefore, if ZES are
the origin of the observed ZBCP, such peak should be present only for the
$d$-wave but not for the $s$-wave material. As shown below, for NCCO indeed no
ZBCP is observed giving strong evidence for the ZES scenario and ruling out
the magnetic interface scattering model.  Our data also show a nonlinear
evolution of the shift of spectral weight to higher energies with increasing
magnetic field.

Up to now in several experiments a ZBCP has been observed in junctions where
only a single electrode was based on a cuprate superconductor
\cite{Alff:97b,Covington:97,Alff:96b,Walsh:92,Lesueur:92,Chen:92,Covington:96}.
Many more experiments with the tunneling direction along the $c$-axis have
been performed, where a ZBCP due to ZES is expected only as an artefact of the
finite surface roughness of the HTS electrode. In our experiments  we used
well defined [001] tilt HTS-GBJs fabricated on bicrystal substrates
\cite{Gross:94a}. It has been shown recently that the quasiparticle transport
mechanism in these junctions is dominated by elastic, resonant tunneling via
localized states making them suitable for spectroscopic studies
\cite{Marx:95,Froehlich:95,Froehlich:97a,Gross:97a}.  A pronounced ZBCP has
been observed in the tunneling conductance of these GBJs which has been
discussed both in terms of ZES \cite{Gross:97a} and the presence of magnetic
scattering centers at the grain boundary \cite{Froehlich:97a}. In this report,
we clearly show that the former analysis can be applied for the HTS-GBJs.
There are several advantages of using GBJs. Firstly, these junctions are
formed by two HTS electrodes and can be fabricated easily from different HTS
materials \cite{Gross:94a,Gross:97a}. Employing {\em in}trinsic interfaces
less problems arise from contamination due to {\em ex-situ} processing of the
samples. Secondly, the tunneling direction for [001] tilt GBJs is along the
$ab$-plane. Thirdly, the direction of tunneling within the $ab$-plane can be
varied by varying the misorientation angle of the bicrystal substrate,
although the faceting of the grain boundary always results in an averaging
over a finite range of angles \cite{Mannhart:96}.  In this context, we note
that an exact quantitative description of effects related to the faceting  is
not yet available.

The GBJs studied in our experiments were prepared on symmetrical [001] tilt
SrTiO$_3$ bicrystals with 24$^{\circ}$ or 36.8$^{\circ}$ misorientation
angles. The fabrication and characterization of the GBJs has been described in
detail elsewhere \cite{Gross:94a,Beck:96a}. The measurements of the
current-voltage ($I(V)$) and conductance vs.~voltage ($G(V)=dI(V)/dV$)
characteristics were performed in a standard four-lead arrangement.

%%%%%%%%%%%%%%%%%%%%%%%%  FIGURE 1 %%%%%%%%%%%%%%%%%%%%%%%%%

\begin{figure}[p]
\noindent
\vspace*{-0.5cm}\\
\hspace*{0cm}\centering\epsfxsize=12.5cm\epsffile{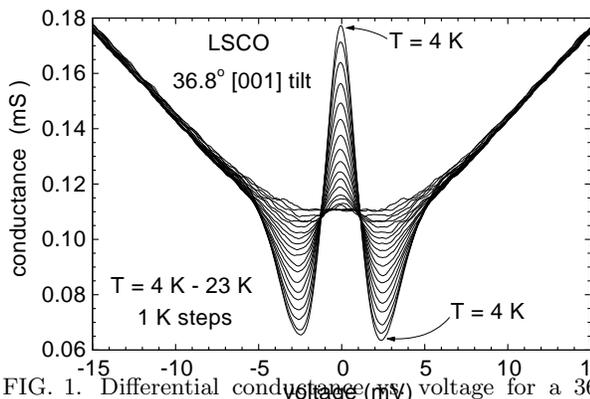}
\vspace*{-3.75cm}
\caption{Differential conductance vs.~voltage for a $36.8^{\circ}$ [001] tilt
LSCO-GBJ between 4 and 23\,K.}
\label{prb97f1}
\end{figure}
\vspace*{-0.4cm}
%%%%%%%%%%%%%%%%%%%%%%%%%%%%%%%%%%%%%%%%%%%%%%%%%%%%%%%%%%%%

\begin{figure}[p]
\noindent
\vspace*{-0.5cm}\\
\hspace*{0cm}\centering\epsfxsize=8.5cm\epsffile{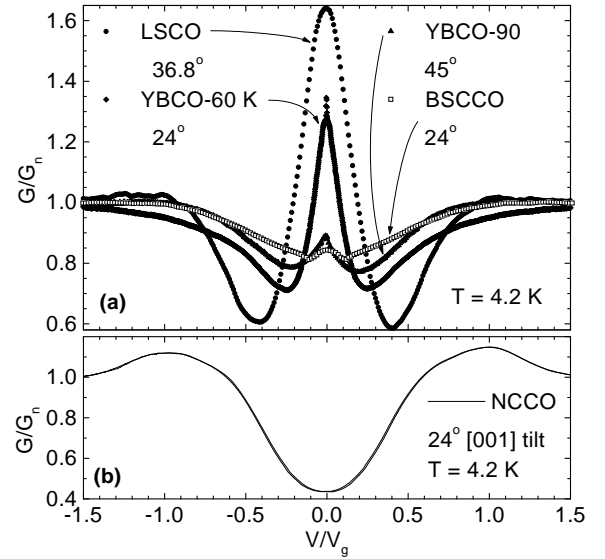}
\vspace*{-4.65cm}
\caption{Normalized conductance vs.~normalized voltage of [001] tilt GBJs
formed by YBCO (90 and 60\ K phase), BSCCO, and LSCO at $T=4.2$\ K. In (b) the
same dependence is shown for a NCCO-GBJ.}
\label{prb97f2}
\end{figure}

Fig.~\ref{prb97f1} shows a set of typical $G(V)$-curves obtained for
LSCO-GBJs. The critical temperature $T_c$ of the LSCO electrodes was about 24\
K. Very similar curves (see Fig.~\ref{prb97f2}) have been measured for oxygen
deficient YBCO ($T_c\approx 60$\ K), BSCCO ($T_c\approx80$\ K)
\cite{Froehlich:97a,Gross:97a}, and fully oxygenated YBCO ($T_c\approx 90$\
K).  At voltages above the gap voltage of the electrode material the
$G(V)$-curves show a temperature independent conductance that has an about
parabolic shape and can be modelled by the influence of the applied voltage on
the shape of the tunneling barrier \cite{Froehlich:97a}. Below the gap voltage
a reduced conductance due to a reduced density of states is observed. With
increasing temperature the conductance increases approaching the normal state
curve $G_n(V)$ with $T$ approaching $T_c$. The height of the ZBCP decreases
with increasing temperature and vanishes at $T=T_c$. For most samples over a
considerable temperature range the decrease follows a $1/T$ dependence. The
temperature evolution of the $G(V)$-curves clearly demonstrates that the
superconducting state is being probed. This proof is important with respect to
the interpretation of the ZBCP in terms of Andreev bound states. The $G(V)$
curves also show that the parabolic background conductance $G_n(V)$ is a
normal state effect, which in the following is eliminated by normalizing
$G(V)$ to $G_n(V)$.

%%%%%%%%%%%%%%%%%%%%%%%%  FIGURE 2 %%%%%%%%%%%%%%%%%%%%%%%%%

\vspace*{-0.4cm}

%%%%%%%%%%%%%%%%%%%%%%%%%%%%%%%%%%%%%%%%%%%%%%%%%%%%%%%%%%%%

In Fig.~\ref{prb97f2} the normalized tunneling conductance, $G/G_n$, of GBJs
fabricated from YBCO, BSCCO, LSCO, and NCCO ($T_c\approx24$\ K) is plotted
versus the voltage normalized to the gap voltage $V_g$. Here, $V_g$
= 25, 20, 15, 6, and 6\ meV was used for BSCCO, YBCO (90\ K phase), YBCO (60\
K phase),  LSCO, and NCCO, respectively. In first approximation $eV_g$ can be
considered to be close to the gap energy $\Delta_0$, in contrast to the
BCS-theory. See for example the calculations in reference \cite{Barash:95a}.
It is evident from Fig.~\ref{prb97f2}a that YBCO, BSCCO and LSCO, for which
the OP is considered to have a dominating $d$-wave component, qualitatively
show the same behavior. A clear gap structure with reduced density of states
is observed in combination with a ZBCP. For BSCCO and YBCO-90 the ZBCP is
reduced in height as compared to YBCO-60 and LSCO. The reason for this
reduction is not clear at present. However, considering the dependence of the
ZBCP on the degree of faceting of the grain boundary, which determines the
amount of averaging over the in-plane crystal directions, this observation is
not surprising. In contrast, for NCCO, which is considered to be a $s$-wave
superconductor, only a gap structure but never a ZBCP is observed as
demonstrated by Fig.~\ref{prb97f2}b.

%%%%%%%%%%%%%%%%%%%%%%%%  FIGURE 3 %%%%%%%%%%%%%%%%%%%%%%%%%

\begin{figure}[p]
\noindent
\vspace*{-0.5cm}\\
\hspace*{0cm}\centering\epsfxsize=8.5cm\epsffile{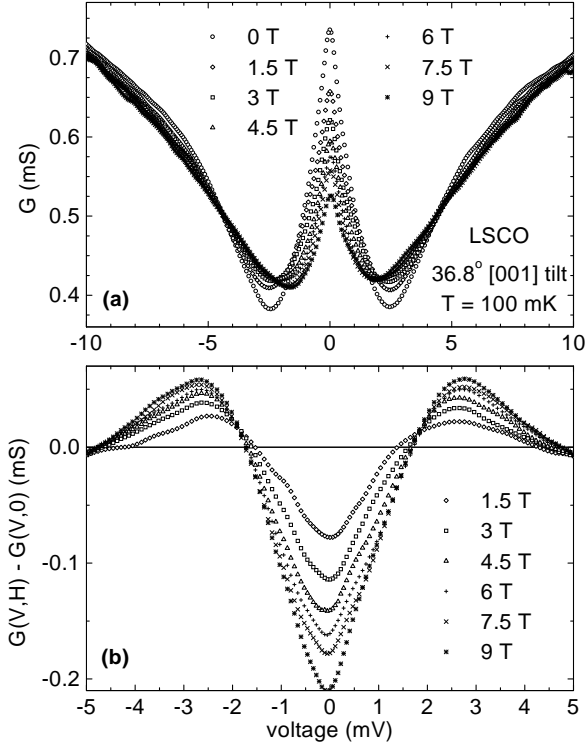}
\vspace*{-2cm}
\caption{(a) Magnetic field dependence of the ZBCP at 100\ mK for a [001] tilt
LSCO-GBJ. In (b), $G(V,H)-G(V,0)$ is plotted for the same sample. The applied
magnetic fields ranged between 1 and 9\ T ($1.5$\ T steps).}
\label{prb97f3}
\end{figure}
\vspace*{-0.4cm}

%%%%%%%%%%%%%%%%%%%%%%%%%%%%%%%%%%%%%%%%%%%%%%%%%%%%%%%%%%%%

\begin{figure}[p]
\noindent
\vspace*{-0.5cm}\\
\hspace*{0cm}\centering\epsfxsize=12.5cm\epsffile{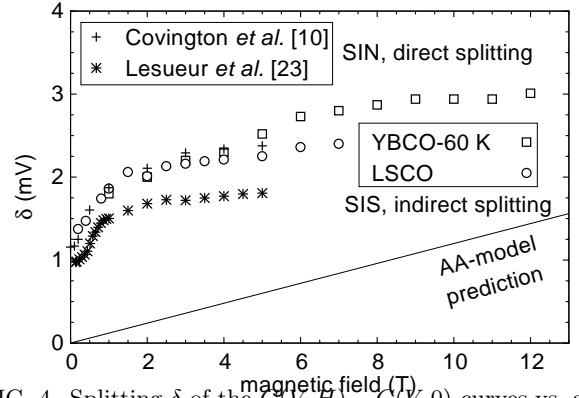}
\vspace*{-3.5cm}
\caption{Splitting $\delta $ of the $G(V,H) - G(V,0)$ curves
vs.~applied magnetic field of a 36.8$^{\circ}$ [001] tilt LSCO-GBJ at $T=100$\
mK and a 24$^{\circ}$ [001] tilt YBCO-GBJ at $T=4.2$\ K. Also shown are data
for the {\em direct} splitting in $G(V,H)$ of SIN-junctions from literature.}
\label{prb97f4}
\end{figure}

We also measured the dependence of the ZBCP on a magnetic field $H$ applied
parallel to the grain boundary plane. A typical result is shown in
Fig.~\ref{prb97f3}. The applied magnetic field reduces the spectral weight at
zero energy and shifts it to finite energies that increase with increasing
field. As shown in Fig.~\ref{prb97f3}b this results in a splitted peak
structure of the difference curve $G(V,H)-G(V,0)$  where the distance between
the peaks is defined as $2\delta$. We emphasize that so far we could not
directly observe the splitting in the $G(V)$-curve down to $T=100$\,mK, where
the thermal smearing amounts to only a few 10\ $\mu$eV. In Fig.~\ref{prb97f4},
$\delta$ is plotted versus $H$ for a LSCO-GBJ at 100\,mK and a YBCO-GBJ at
$4.2$\ K together with data of the $direct$ split in $G(H)$ taken from
literature. Clearly, $\delta$ does not vary linearly with $H$ as predicted by
the AA-model. For all investigated samples $\delta$ increases slower than
linearly and tends to saturate at high fields in agreement with results
published recently \cite{Covington:97}.

%%%%%%%%%%%%%%%%%%%%%%%% FIGURE 4 %%%%%%%%%%%%%%%%%%%%%%%%%

\vspace*{-0.4cm}

%%%%%%%%%%%%%%%%%%%%%%%%%%%%%%%%%%%%%%%%%%%%%%%%%%%%%%%%%%%

We first will discuss our experimental findings in terms of the AA-model
\cite{Appelbaum:66a}. For tunneling across a barrier containing localized
spins beyond a contribution $G_1$ due to direct tunneling without interaction
with the spins there are two further contributions $G_2$ and $G_3$ to the
total conductivity. The first ($G_2$) is related to tunneling involving a spin
exchange and the latter ($G_3$) to Kondo-type scattering. According to the
AA-model \cite{Appelbaum:66a} one expects $G_3(V,T) \propto {\rm ln}
[E_0/(|eV| + k_BT)]$, where $E_0$ is a cut-off energy. Applying a magnetic
field the Zeemann splitting of the impurity levels causes a dip of width
$2\delta =2g\mu _BH$ due to a reduction of $G_2$. Here, $g$ is the $g$-factor
and $\mu _B$ the Bohr magneton. Furthermore, in an applied field the Kondo
peak is split into three peaks separated by $\delta$ with the zero bias peak
completely suppressed.  This results in a peak of $G(V,H) - G(V,0)$ at $eV=\pm
\delta $. The AA-model predicts a peak-to-peak width $2\delta =2g\mu
_BH$ that increases $\propto H$, i.e. $\delta/H = g\mu _B \simeq 0.12$\ meV/T
for $g\simeq 2$. This is in clear contradiction to our results, which show
both much larger absolute values of $\delta/H$ up to more than 2\,meV/T and a
strong increase of $\delta/H$ with decreasing $H$, in agreement with other
data reported in literature \cite{Covington:97,Lesueur:92}. Furthermore, for
all YBCO, BSCCO and LSCO samples the ZBCP always disappeared just at $T_c$,
which is significantly different for the different materials. This is very
difficult to be explained within the AA-model, which predicts the ZBCP to
decrease with increasing $T$ but not to disappear at a specific temperature.
Finally, within the AA-model the absence of the ZBCP for NCCO would imply the
absence of magnetic scatterers for this material. Supposing that magnetic
scatterers at grain boundaries result from oxygen loss and the formation of
magnetic Cu$^{2+}$-ions, the basic difference between NCCO and the other
materials is difficult to understand.

We now turn to the Andreev bound state model. As discussed above, in this
model the ZBCP arises from bound states formed by the constructive
interference of quasiparticles that propagate through different order
parameter fields incident and reflecting from the surface of the junction
electrode.  ZES are formed if the scattering induces a change in sign of the
OP. Hence, ZES are not possible for a $s$-wave symmetry of the OP. However, in
the case of a dominating $d_{x^2-y^2}$-symmetry of the OP, at all surfaces
parallel to the $c$-axis ZES are formed except for the surfaces exactly
perpendicular to the $a$- or $b$-axis direction. Hence, the ZES-model
naturally accounts for the observation that a ZBCP is observed only for YBCO,
BSCCO and LSCO, which most likely have a dominating $d$-wave component of the
OP, whereas it is absent for NCCO, which is supposed to have a dominating
$s$-wave OP. The ZES-model also qualitatively accounts for the increase of the
height of the ZBCP with decreasing temperature and the nonlinear shift of the
peak spectral weight to finite voltages with increasing magnetic field
\cite{Barash:97,Fogelstroem:97a,Kashiwaya:95a,Fogelstroem:98}. For example,
for a surface to $a$-axis orientation of 20$^{o}$, $G(0,T)/G_n(0)$ was
predicted to decrease about $\propto 1/T$ \cite{Barash:97} in fair agreement
with our data. A detailed quantitative analysis of our experimental data still
is not possible, since no prediction of the exact $T$ and $H$ dependence of
the ZBCP is available taking into account the angle averaging due to the
faceting of the grain boundaries.

We finally would like to address the possibility of a surface state with
broken time reversal symmetry as predicted by Fogelstr\"{o}m et {\it al.}
\cite{Fogelstroem:97a} and experimentally observed by Covington et {\it al.}
\cite{Covington:97}. In this case the ZBCP is expected to split in zero
magnetic field. Such splitting has not been observed directly in our
experiments down to temperatures of 100\ mK for LSCO and 4.2\ K for YBCO
similar to other experiments \cite{Ekin:97}. A possible reason for this
observation may be the considerable faceting of the grain boundary plane
together with impurity scattering that suppresses the field splitting of the
ZBCP \cite{Fogelstroem:98}. This is the reason why the observed behavior of
$\delta$ vs.~$H$ does not provide definitive evidence for a subdominant
$s$-wave OP and time-reversal symmetry breaking at the grain boundary
interface. This issue has to be clarified by future experimental and
theoretical work taking into account the grain boundary faceting and impurity
scattering.

In conclusion, it has been shown that quasiparticle tunneling in GBJs can be
used for probing the symmetry of the order parameter in HTS. The tunneling
spectra of [001] tilt GBJs formed by YBCO, BSCCO and LSCO were found to always
show a ZBCP while such peak is absent for NCCO. The height of the ZBCP
decreases with increasing temperature and disappears at $T_c$. These
observations are not compatible with the assumption of tunneling involving
magnetic impurities as described by the AA-model, but can naturally be
explained by the presence of zero energy Andreev bound states at surfaces of
HTS. The existence of ZES represents a further proof that the order parameter
of YBCO, BSCCO and LSCO changes sign on the Fermi surface and most likely has
a dominating $d$-wave component. The tunneling data of NCCO is consistent with
an anisotropic $s$-wave symmetry of the pair potential in the electron doped
HTS. The evolution of the ZBCP with varying applied magnetic field and
temperature can be qualitatively described within the $d$-wave scenario.

We wish to thank H.~Burkhardt, J.~Halbritter, S.~Kashiwaya, D.~Rainer,
S.~Scheidl, and Y.~Tanaka for stimulating discussions, and M.~Fogelstr\"{o}m
and J.~A.~Sauls for performing calculations on the influence of disorder on
the magnetic field behavior, and M.~Covington for cooperating. This work is
supported by the Deutsche Forschungsgemeinschaft (SFB 341).

\end{multicols}

\end{document}